# A Novel Multiple Interval Prediction Method for Electricity Prices based on Scenarios Generation: Definition and Method

Xin Lu

**Abstract:** This paper presents interval prediction methodology to address limitations in existing evaluation indicators and improve prediction accuracy and reliability. First, new evaluation indicators are proposed to comprehensively assess interval prediction methods, considering both all-sample and single-sample scenarios. Second, a novel Pattern-Diversity Conditional Time-Series Generative Adversarial Network (PDCTSGAN) is introduced to generate realistic scenarios, enabling a new interval prediction approach based on scenario generation. The PDCTSGAN model innovatively incorporates modifications to random noise inputs, allowing the generation of pattern-diverse realistic scenarios. These scenarios are further utilized to construct multiple interval patterns with high coverage probability and low average width. The effectiveness of the proposed methodology is demonstrated through comprehensive case studies. The paper concludes by highlighting future research directions to further enhance interval prediction methods.

1. Introduction

　　Accurately forecasting electricity prices has garnered significant attention in recent years [1, 2], as it plays a crucial role [3, 4] in guiding market participants in areas such as profit maximization, risk management, and ensuring stable operations. Electricity prices are influenced by numerous factors and are often characterized as anti-persistent data. These factors significantly impact electricity demand and renewable energy generation, contributing to price fluctuations and presenting notable challenges for electricity price forecasting. Electricity price prediction can be broadly divided into point prediction and interval prediction. Point prediction involves deterministic methods that generate specific predicted values by minimizing metrics such as root mean square error (RMSE) and mean absolute error (MAE) between forecasts and actual outcomes. However, uncertainties in electricity markets, often stemming from incomplete data or unforeseen events, limit the effectiveness of point prediction methods. To address these uncertainties, interval prediction methods have been developed, offering a way to account for variability and mitigate the inaccuracies inherent in point predictions.

　　The base model of this study is CTSGAN, which was published in [5]. After further improvements, this work has been published in [6, 7]. We applied this prediction model to the fields of batteries [8], virtual power plants [9, 10], and shared energy storage [11], achieving promising results in each application [12].

This paper makes the following contributions:

1. New evaluation indicators based on all samples and one sample are proposed to address the fact that the evaluation indicators currently used for interval prediction do not adequately reflect the strengths and weaknesses of the prediction methods.
2. A novel pattern-diversity conditional time-series GAN (PDCTSGAN) is proposed, which can generate realistic scenarios, and a novel interval prediction method is proposed based on scenario generation.
3. By modifying the random noise input, the PDCTSGAN model can generate pattern-diversity realistic scenarios, with which can be combined multiple pattern-diversity intervals with high coverage probability and low average width.

The rest of this paper is organized as follows: Section 2 introduces the definition and evaluation of prediction intervals. Section 3 illustrates the steps of intervals prediction and proposes the methodology of the PDCTSGAN model. Section 4 conducts the case studies, followed by the conclusion and some discussions on future research directions.

2. Definition and Evaluation of prediction intervals

2.1 Reliability and sharpness of prediction intervals

The reliability and sharpness are two indispensable indicators of prediction intervals [13]. The empirical coverage probability of all samples (ECPAS) and empirical average width of all prediction intervals (EAWAPI) are used to evaluate the reliability and sharpness of prediction intervals for all samples $\boldsymbol{\theta} = [\theta_1, \cdots \theta_t, \cdots \theta_T]$, respectively, as follows:

$$\delta = \frac{1}{T}\sum_{t=1}^{T} c_t, \quad \begin{cases} c_t = 1, & \text{if } \theta_t \in [L_t, U_t] \\ c_t = 0, & \text{if } \theta_t \notin [L_t, U_t] \end{cases} \tag{1}$$

$$\xi = \frac{1}{T}\sum_{t=1}^{T}(U_t - L_t) \tag{2}$$

where $\delta$ represents ECPAS and $\xi$ represents EAWAPI; T represents the size of the samples and t represents the current number of all samples; $\theta_t$ denotes the $t^{th}$ sample of all samples, and $L_t$ and $U_t$ denote the predicted lower bound and upper bound. $c_t$ is the indicator that equal to 1 if prediction interval $[L_t, U_t]$ covers $\theta_t$, and to 0 otherwise [14].

2.2 ECPAS and EAWAPI of all samples and confidence level

In [13, 15-17], the ECPAS $\delta$ and EAWAPI $\xi$ are the target of prediction intervals with neural networks. However, these methods usually optimize the weights of a particular neural network using a heuristic algorithm which is not guaranteed to find the global optimum every time and often falls into a local optimum, in which case the intervals generated by the neural network is not optimal. Therefore, there is a bias in using the $\delta$ and $\xi$ of a single interval prediction alone to access the validity of the prediction method. Multiple prediction intervals can be obtained through repeated simulations and confidence levels can be introduced to evaluate multiple prediction intervals based on a particular method.

The definition of confidence interval by Neyman [18] in 1937 as the following:

**Definition 1** (Confidence interval for one parameter) *An X% confidence interval for a parameter θ is an interval (L,U) generated by a procedure that in repeated sampling has an X% probability of containing the true value of θ, for all possible values of θ*[19].

Confidence level is often used to estimate the minimum certainty that the claimed condition is accurate. The condition here could be a larger than or equal to Y% ECPAS, or even a smaller than or equal to EAWAPI Z. According to the Definition 1, the definitions of confidence level for ECPAS and EAWAPI are first proposed as follows:

**Definition 2** (Confidence level for a Y% ECPAS) *An X% confidence level for a Y% ECPAS is an ECPAS calculated by prediction intervals generated by a procedure that in repeated sampling has an X% probability of being larger than or equal to Y%, for all possible ECPAS.*

**Definition 3** (Confidence level for an EAWAPI Z) *An X% confidence level for EAWAPI Z is an EAWAPI calculated by prediction intervals generated by a procedure that in repeated sampling has an X% probability of being smaller than or equal to Z, for all possible EAWAPI.*

With above Definition 2 and Definition 3, the confidence level for ECPAS $\phi$ are given as follows:

$$\delta^s = \frac{1}{T}\sum_{t=1}^{T} c_t, \quad \begin{cases} c_t = 1, if\ \theta_t \in [L_t^s, U_t^s] \\ c_t = 0, if\ \theta_t \notin [L_t^s, U_t^s] \end{cases} \tag{3}$$

$$\phi = \frac{1}{S}\sum_{s=1}^{S} b^s, \quad \begin{cases} b^s = 1, if\ \delta^s \geq \delta' \\ b^s = 0, if\ \delta^s < \delta' \end{cases} \tag{4}$$

where S represents the total number of repeated prediction process, and s represents the s$^{th}$ prediction process; $\delta^s$ is different from $\delta$ in (1), which denotes the s$^{th}$ ECPAS obtained from s$^{th}$ prediction result; $L_t^s$ and $U_t^s$ denote the predicted lower bound and upper bound from s$^{th}$ repeated predication process for the same sample $\theta_t$, respectively; And $b^s$ denote the indicator, $b^s$ equal to 1 if $\delta^s$ is equal to or larger than a given ECPAS with $\delta'$, and otherwise to 0. And other parameters have the same meaning as (1) and (2). In (4), the given ECPAS with $\delta'$ can be fixed, the greater the confidence level $\phi$, the better; or the confidence level with $\phi$ can be fixed, the greater the ECPAS $\delta'$, the better.

The following figure gives a simple example of the difference between a Y% ECPAS ($\delta^s$=Y%) and confidence level for an ECPAS Y% ($\phi$=Y%).

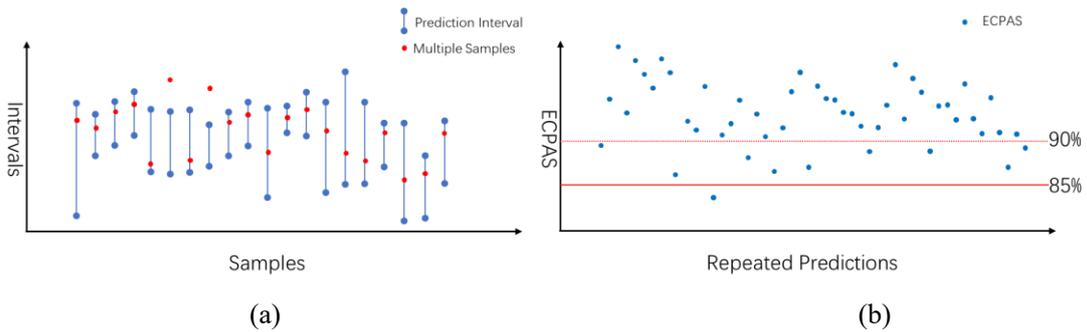

(a)          (b)

Fig. 1. Difference between (a) ECPAS and (b) confidence level for ECPAS

Fig.1 (a) gives the empirical coverage probability of all samples. There are a total of 20 actual samples, and 2 samples are not covered by the prediction intervals, so the ECPAS is 90%. This prediction process is repeated 50 times, and the ECPAS is different each time as Fig. 1 (b). The ECPAS being larger than or equal to 90% (above the dotted red line) occurs 40 times out of 50, which is known as a 90% confidence level. This situation can be said that the ECPAS will be larger than or equal to 90% with a confidence level of 90%, or a 90% confidence level for ECPAS will be equal to or larger than 90%. By changing ECPAS given value $\delta'$, another conclusion can be obtained, a 95% confidence level for ECPAS will be equal to or larger than 85% (above the solid red line).

The confidence level for EAWAPI $\varphi$ are given as follows:

$$\xi^s = \frac{1}{T}\sum_{t=1}^{T}\left(U_t^s - L_t^s\right) \tag{5}$$

$$\varphi = \frac{1}{S}\sum_{s=1}^{S}a^s, \quad \begin{cases} a^s = 1, & \text{if } \xi^s < \xi' \\ a^s = 0, & \text{if } \xi^s \geq \xi' \end{cases} \tag{6}$$

where $\xi^s$ is different from $\xi$ in (2), which denotes the s$^{th}$ EAWAPI obtained from s$^{th}$ prediction results; $a^s$ denote the indicator, $a^s$ equal to 1 if $\xi^s$ is equal to or smaller than a given EAWAPI with $\xi'$, and otherwise to 0. And other parameters have the same meaning as (1) and (2). In (6), the given EAWAPI with $\xi'$ can be fixed, the greater the confidence level $\varphi$, the better; or the confidence level with $\varphi$ can be fixed, the smaller the EAWAPI $\xi'$, the better.

2.3 ECP and EAW of one sample

With the current developments in energy management, the demands on prediction are increasing and the evaluation of the whole samples cannot meet the requirements of forecast data users, especially for some uncertainty prediction for spikes and troughs, which are becoming increasingly important. The ECPAS and EAWAPI in section 2.1 and 2.2 are introduced to evaluate the intervals for all the test samples. It is not sufficient to evaluate the whole samples alone; the poor sample in the prediction need to be evaluated individually. Empirical coverage probability (ECP) and empirical average width (EAW) of prediction intervals for one sample $\theta_t$ are given as follows:

$$\delta_t = \frac{1}{S}\sum_{s=1}^{S}c_s, \quad \begin{cases} c_s = 1, & \text{if } \theta_t \in [L_t^s, U_t^s] \\ c_s = 0, & \text{if } \theta_t \notin [L_t^s, U_t^s] \end{cases} \tag{7}$$

where $\delta_t$ represents the ECP for one sample $\theta_t$ obtained by a total of S repeated predictions; $c_s$ is the indicator that equal to 1 if s$^{th}$ prediction interval $[L_t^s, U_t^s]$ covers $\theta_t$, and to 0 otherwise.

$$\xi_t = \frac{1}{S}\sum_{s=1}^{S}\left(U_t^s - L_t^s\right) \tag{8}$$

where $\xi_t$ represents the EAW for one sample $\theta_t$ obtained by a total of S repeated predictions.

3. Prediction steps and methodology

3.1 Novel intervals prediction based on scenarios generation

The conventional intervals prediction methods are generally divided into two kinds, one is to estimate the mean and standard deviation separately, and finally combine them, such as MVE method. The second is to directly give the upper and lower bounds of the intervals, such as LUBE and Bootstrap method. The ideal prediction intervals need to contain as many scenarios as possible with different probabilities. If all possible scenarios can be generated, then these scenarios can be combined to obtain an interval. This paper proposes a novel interval prediction method based on scenarios generation as Fig. (2).

The common neural networks cannot generate a large number of different scenarios, because the network parameters are fixed after the training is completed, and there is a one-to-one mapping relationship between input and output.

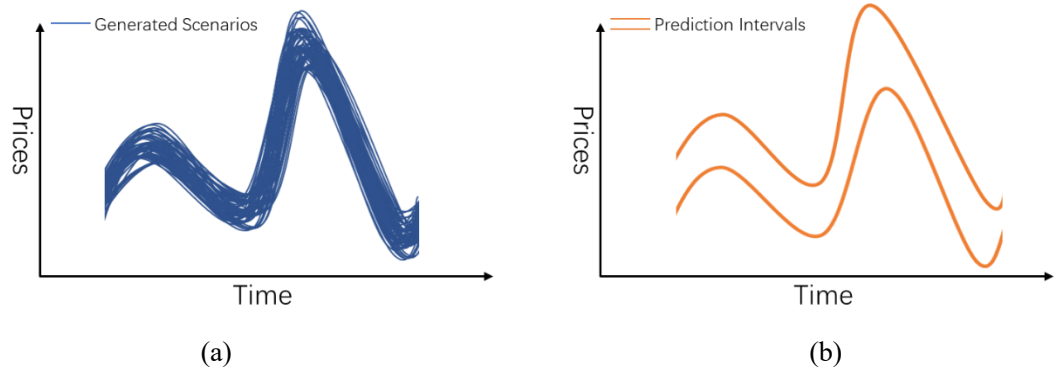

(a) (b)

Fig. 2. Prediction Intervals based on scenarios generation (a) Generated Scenarios and (b) Prediction Intervals

3.2 Intervals prediction steps

The interval prediction flowchart is as Fig. 4.

1) Data collection and pre-processing. Data can be collected from conducted on a publicly database. For example, Australian electricity price data and electricity demand data can be downloaded from the AEMO website. Ref. [20] pointed out that outliers can affect the accuracy of prediction results, and it is necessary to limit the data to a certain range. In addition, for making the dataset fit the prediction model better, the data should be normalized to the range [0,1] with a min-max normalization method as follows:

$$p' = \frac{p - p_{\min}}{p_{\max} - p_{\min}} \quad (9)$$

where $p_{min}$ and $p_{max}$ are the maximum and minimum value in dataset, respectively. $p$ and $p'$ are the original value and normalized value, respectively.

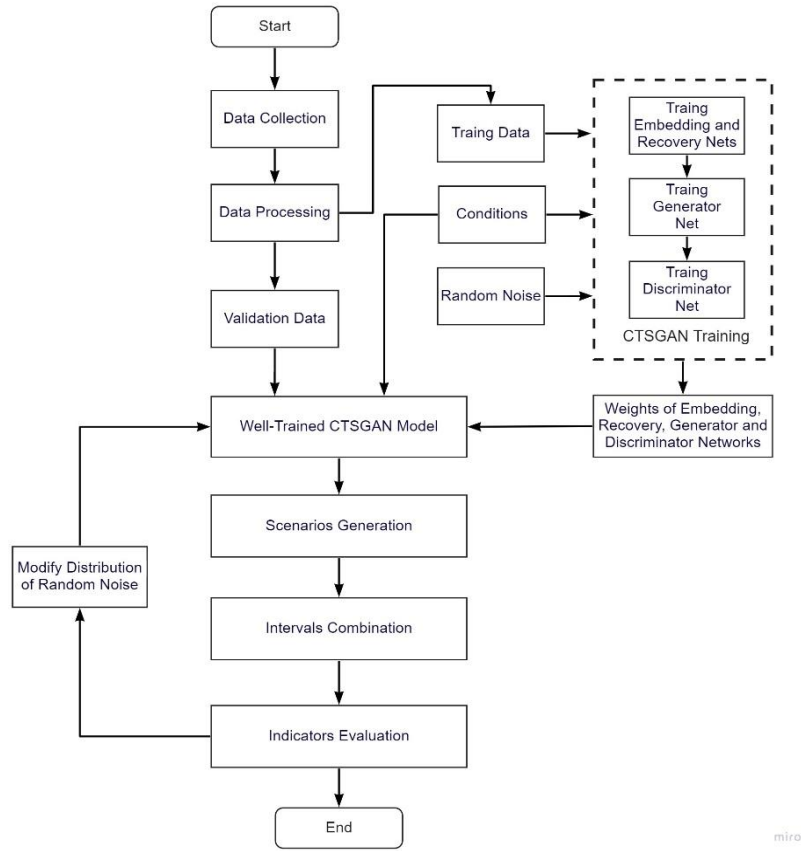

Fig. 4 Flowchart of interval prediction

2) CTSGAN weights optimization. The inputs of the CTSGAN are the pre-processing data, random noise, and conditions. The training process can be divided into three stages. In the first stage, the optimization objectives are the weights of the embedding and recovery networks. In the second stage, the optimization objective is the weights of the generator. The generator receives the mapped data computed by embedding network and generates the next latent vector. The final stage is the joint training stage, in which the optimization objectives are the weights of the generator and the discriminator.

3) Scenarios generation and interval prediction. After CTSGAN model is well trained, it can be used to predict next scenario. Firstly, with random noise as input, different scenarios can be obtained and be combined a prediction interval. Secondly, repeat scenarios generation step and multiple prediction intervals can be obtained. Finally, evaluation indicators of multiple intervals can be calculated.

4) Low-probability scenarios reinforcement prediction. By introducing a variety of different distribution random noise input, CTSGAN could generate pattern-diversity scenarios of low probability scenarios. With pattern-diversity scenarios, the prediction intervals could be more reasonable and have better evaluation indicators.

### 3.3 Conditional time series GAN

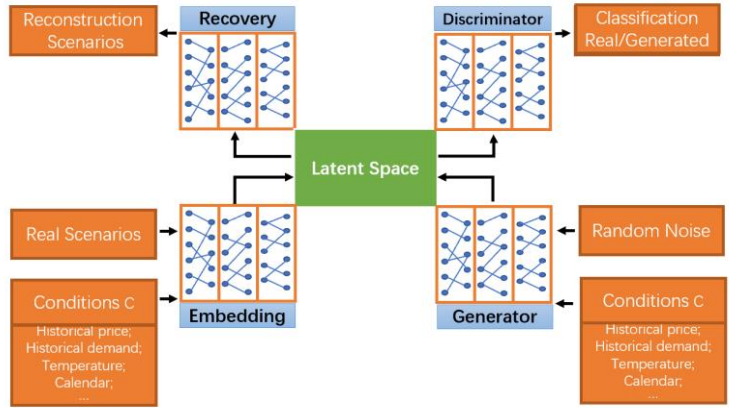

Fig.5. Structure of the CTSGAN model

The novel intervals prediction method is based on scenarios generation, which means that which means that the number of generated scenarios needs to be large enough and relatively realistic. Common neural networks cannot generate a large number of different scenarios, because the network parameters are fixed after the training is completed, and there is a one-to-one mapping relationship between input and output. GAN is introduced here to generate different scenarios, in which the introduction of random noise as input makes the generated scenarios diversified. On the other hand, the authenticity of the generated scenarios is also very important for the interval construction. The conventional GAN is an unsupervised learning algorithm, so the conventional GAN needs to be modified to a supervised algorithm to make the generated scenarios realistic. Inspired by Yoon's work[21], combining the TSGAN and conditions, CTSGAN is proposed to generate to generate a large number of realistic and diversity-rich scenarios, which can be used as the basis for intervals construction.

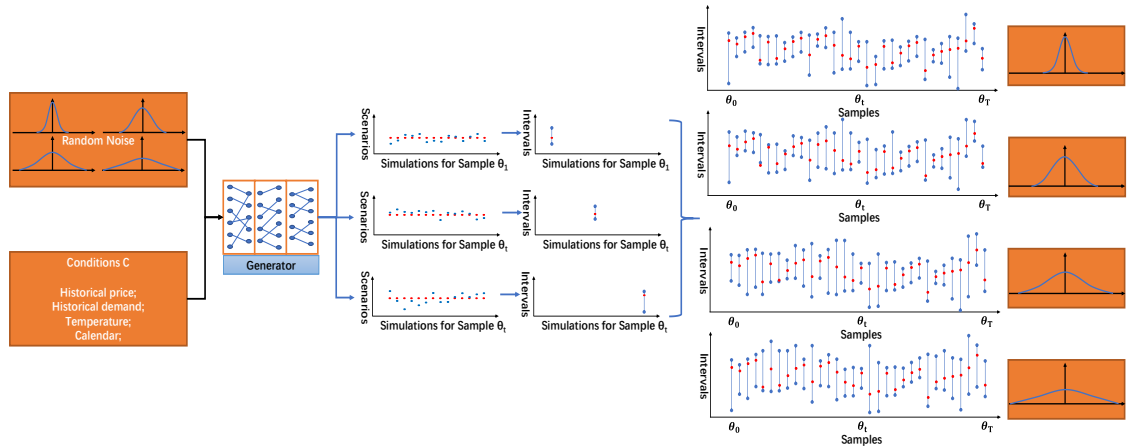

Fig. 6 Pattern- diversity prediction with well-trained CTSGAN

### 3.4 Pattern-diversity CTSGAN and low-probability scenarios reinforcement prediction

In the CTSGAN training and prediction process, random noise $z$ that satisfies the Gaussian distribution $\mathcal{N}(\mu,\sigma)$ is often selected as the input, and the value of $\sigma$ is usually set to 1 and $\mu$ to 0, which is known as the standard Gaussian distribution function. For the standard Gaussian distribution, the random noise $z$ less than $1\sigma$ away from $\mu$ accounts for 68.27% of the set; while $2\sigma$ away from $\mu$ accounts for 95.45%; and $3\sigma$ accounts for 99.73%. This means that the random noise $z$ larger than $1\sigma$ away from $\mu$ accounts for 0.27% of the set. The training process in CTSGAN has more than one million iterations, and low-probability noise, such as random noise $z$ larger than $2\sigma$ or $3\sigma$ away from $\mu$, can always be selected to participate in training. However, in the prediction process, the numbers of prediction of a point will not reach the order of one million, which will cause low-probability random noise to not participate in the prediction, resulting in poor prediction of some low-probability scenarios, like spikes.

There are two main approaches to address the poor diversity generation problem in GANs. One is focus on discriminators by increasing different divergence metrics[22] or optimization the training process[23]. Another is introducing the auxiliary networks[24] or encoders[25]. Some recent research focus on the relationship between random noise input and images generation, such as random noise size[26] and space distance[27]. Above methods are often used in unsupervised GANs to generate rich diversity images and hard to be combined to supervised CTSGAN for low-probability scenarios prediction.

Inspired by above work, a novel PDCTSGAN prediction method is first proposed and introduced to solve poor low-probability scenarios problems, as shown in Fig. 6. After CTSGAN model well- trained with random noise which satisfies the standard Gaussian distribution, different random noise satisfying Gaussian distribution with different standard deviation $\sigma$ is selected as input, and CTSGAN can generate more diverse scenarios and the combined prediction intervals could be rich diversity and contain more low probability scenarios.